# *Stability Analysis of Path-vector Routing*

D.Papadimitriou[(1)] and A.Cabellos[(2)]

[(1)] *Alcatel-Lucent Bell Labs, Antwerpen, Belgium*
[(2)] *Technical University of Catalonia, Barcelona, Spain*

**Abstract**: Most studies on path-vector routing stability have been conducted empirically by means of ad-hoc analysis of BGP data traces. None of them consider prior specification of an analytic method including the use of stability measurement metrics for the systematic analysis of BGP traces and associated meta-processing for determining the local state of the routing system. In this paper, we define a set of metrics that characterize the local stability properties of path-vector routing such as BGP (Border Gateway Protocol). By means of these stability metrics, we propose a method to analyze the effects of BGP policy- and protocol-induced instability on local routers.

## 1 Introduction

Research efforts to understand BGP's instability led to classify them as policy- or protocol-induced to account for the distinction between protocol operation and the inherent behavior of the underlying path-vector routing algorithm: 1) *Policy-induced instabilities*: addressing routing stability consistently with planned BGP routing policy implies eliminating non-deterministic routing states resulting from policy interactions and in particular, non-deterministic and unintended but unstable states. Griffin et al.'s seminal work [1] modeled BGP as a distributed algorithm for solving the Stable Paths Problem, and derived a general sufficient condition for BGP stability, known as "No Dispute Wheel". This sufficient condition guarantees the existence of a stable solution to which BGP always converges. Informally, this sufficient condition allows nodes to have more expressive and realistic preferences than always preferring shorter routes to longer ones. The game theoretic approach introduced in [2] relies on the best-reply BGP dynamics: a convergence game model in which each Autonomous System (AS) is instructed to continuously execute the following actions: i) receive update messages from BGP peering nodes announcing their routes to the destination, ii) choose a single peering node whose route is most preferred to send traffic to, iii) announce the new route to peering nodes. However, as proved in [2], best-reply BGP dynamics is not incentive-compatible even if No Dispute Wheel condition holds: even if all but one AS are following the BGP rules, the remaining AS may not have the incentive to follow them. Interestingly, as demonstrated in [2], incentive compatibility of best-reply BGP dynamics requires combining an additional global condition (Route Verification) together with the "No Dispute Wheels" to guarantee stability. Consequently, all known conditions for global stability are sufficient but not necessary conditions (checking them is an NP-hard problem and enforcing them requires a global deployment of an additional mechanism); on the other hand, local instability effects have yet to be characterized; 2) *Protocol-induced instabilities*: BGP is an inter-AS path-vector routing protocol subject to Path Exploration phenomenon like any other path-vector algorithm. Indeed, BGP routers may announce as valid, routes that are affected by a topological change and that will be withdrawn shortly after subsequent routing updates. This phenomenon is the main reason for the large number of routing updates received by BGP routers which exacerbate inter-domain routing system instability and processing overhead [3]. Both result in delaying BGP convergence time upon topology change/failure [4]. Several mitigation mechanisms exist to partially limit the effects of path exploration; however, none actually eliminate its effects. Hence, BGP is intrinsically subject to instability.

The objectives for investigating path-vector routing stability are to 1) Develop a method to process and interpret the data part of BGP routing information bases in order to identify and characterize occurrences of BGP routing system instability; such characterization can be used as a comparison point for the stability of the newly developed schemes (candidate to replace BGP) and characterize instability phenomena any routing system would have to cope with; 2) Determine a set of stability metrics and develop methods for using them in order to provide a better understanding of the BGP routing system's stability; 3) Investigate how path-vector routing behavior and network dynamics mutually influence each other. The proposed method aims to bring rigor and consistency to the study of routing stability.

## 2 Routing Stability and Metrics

### *2.1 Preliminaries*

The AS topology of the routing system is described as a graph $G = (V,E)$, where the vertices (nodes) set $V$, $|V| = n$, represents the AS, and the edges set $E$, $|E| = m$, represents the links between AS. At each node $u \in V$, a route $r$ per destination $d$ ($d \in D$) is selected and stored as an entry in the local routing table (RT) whose total number of entries is denoted by $N$, i.e., $|RT| = N$. At node $u$, a route $r_i$ to destination $d$ at time $t$ is defined by $r_i(t) = \{d, (v_k=u, v_{k-1},…,v_0=v), A\}$ with $k > 0 \mid \forall\, j, k \geq j > 0$, $\{v_j, v_{j-1}\} \in E$ and $i \in [1,N]$, where $(v_k=u, v_{k-1},…,v_0=v)$ represents the AS-path, $v_{k-1}$ the next hop of $v$ along the AS-path from node $u$ to $v$, and $A$ its attribute set. Let $P_{(u,v),d}$ denote the set of paths from node $u$ to $v$ towards destination $d$ where each path $p(u,v)$ is of the form $\{(v_k=u, v_{k-1},…,v_0=v), A\}$. A routing update leads to a change of the AS-path $(v_k, v_{k-1},…,v_0)$ or an element of its attribute set $A$. A withdrawal is denoted by an empty AS-path ($\varepsilon$) and $A = \varnothing$: $\{d,\varepsilon,\varnothing\}$. According to the above definition, if there is more than one AS-path per destination $d$, they will be considered as multiple distinct routes.

## 2.2 Stability definitions

The stability of a routing system is characterized by its response (in terms of processing of routing information) to inputs of finite amplitude. Routing system inputs may be classified as i) internal system events such as routing protocol configuration change or ii) external events such as those resulting from topological changes. Both types of events lead to the exchange of routing updates that may result in routing states changes. Indeed, BGP (and in general any path-vector routing) does not differentiate routing updates with respect to their root cause, their identification (origin), etc. during its selection process.

*Definition 1*: Let RT(t) represent the routing table at some time t. At time t+1, RT(t+1) = $RT_0(t) \oplus \Delta RT(t+1)$ where, $RT_0(t)$ is the set of routes that experience no change between time t and t+1, and $\Delta RT(t+1)$ accounts for all route changes (additions, deletions, and changes to previously existing routes) between time t and t+1.

The magnitude of the output of a stable routing system should be small whenever the input is small. That is, a single routing information update shall not result in output amplification. Equivalently, a stable system's output will always decrease to zero whenever the input events stop. A routing system, which remains in an unending condition of transition from one state to another when disturbed by an external or internal event, is considered to be unstable. Provide means for measuring the magnitude of the output is the main purpose of the metric referred to as "stability of the selected route". For this purpose, we define the criteria for qualifying the effect of a perturbation on the local routing table so as to locally characterize the stability of the routing system. More precisely, let $|\Delta RT(t+1)|$ be the magnitude of the change to the routing table (RT) between time t = $t_0$ + k to t + 1 = $t_0$ + (k +1), where $t_0$ is the starting time of the measurement sequence, we distinguish three different equilibrium states for the routing table:

*Definition 2*: when disturbed by an external and/or internal event, a RT is considered to be *stable* if the following condition is met: $|\Delta RT(t+1)| \leq \alpha$, $t \to \infty$, where $\alpha > 0$ is small. In these conditions, if the routing system returns locally to its initial equilibrium state, it is considered to be (asymptotically) stable.

*Definition 3*: when disturbed by an external and/or internal event, a RT is considered to be *marginally stable* if the following condition is met: $\alpha < |\Delta RT(t+1)| \leq \beta$, $t \to \infty$, where $\beta > 0$ is small, $\alpha < \beta$. In these conditions, if the routing system transitions locally to a new equilibrium state, it is considered to be marginally stable.

*Definition 4*: when disturbed by an external and/or internal event, a RT is considered to be *unstable* if the following condition is met: $|\Delta RT(t+1)| > \beta$, $t \to \infty$. In these conditions, the routing system remains locally in an unending condition of transition from one state to another and it is considered to be locally unstable

The values $\alpha$ and $\beta$ shall be set based on operational criteria. Among other factors, $\alpha$ and $\beta$ depend on the observation sampling period that must be set to the Minimum Routing Advertisement Interval (MRAI) in order to ensure one routing update per sampling period. A similar reasoning to the one applied for the Loc_RIB stability (that corresponds to the BGP routing table) can be applied to the Adj_RIB_In (which stores incoming routes from neighbors). It is also interesting to measure the instability induced by the BGP selection process.

## 2.3 Stability metrics

To measure the degree of stability of the Loc_RIB, Adj_RIB_In, and determine how close the routing system is to being unstable the following stability metrics are defined.

The *stability* $\varphi_i(t)$ of selected routes $r_i(t)$ characterizes the stability of the routes $r_i$ (i $\in$ [1,|D|]) stored at time t in the Loc_RIB (|Loc_RIB| = N) by quantifying the magnitude of change for these routes from time t to t+1. This metric quantifies the magnitude of change for these routes between time t = $t_0$+k to t+1 = $t_0$+(k+1), where $t_0$ is the starting time of the measurement sequence (time units are counted by default in terms of minimum routing advertisement interval (MRAI)), and the integer k accounts for the number of MRAI times that have elapsed since the starting time of the measure sequence. The latter determines the minimum amount of time that must elapse between a routing advertisement of a route to a particular destination by a BGP peer. This metric quantifies thus the magnitude of change to route $r_i$, and a routing table with a periodicity determined by the MRAI time. This metric can be directly computed by using the algorithm described in Fig.1.

```
When route r_i is created: φ_i(t) ← 0
if r_i experiences a path or an attribute change (r_i(t+1) ≠ r_i(t)) then φ_i(t+1) ← φ_i(t) + 1
else /* r_i experiences no changes */
    if φ_i(t) = 0 then φ_i(t+1) ← 0
    else if φ_i(t) > 0 then φ_i(t+1) ← φ_i(t) - 1
        end if
    end if
end if
```

**Fig. 1** – *Stability of individual routes*

The computation of the stability metric for an entire routing table (RT) can then be derived from the stability of its individual routes (see Fig.2). Let $|\Delta r_i(t+1)|$ denote the change in stability metric for a single route $r_i$ from time t to t+1. These values are used to compute $|\Delta RT(t+1)|$ defined as the change in stability metric for the entire routing table from time t to t+1. Moreover, $|\Delta RT(t+1)|$ is normalized so that $0 \leq |\Delta RT(t+1)| \leq 1$, where 0 implies perfect stability, and 1 indicates complete instability.

The *most stable route* in the Adj_RIB_In (|Adj_RIB_In| = M) quantifies the relative stability between incoming routes to the same destination d as learned from all upstream BGP peers (i.e., downstream from the point of view of the AS-path towards

destination d) and the one amongst them determined at time t as the most stable. For this purpose, let $W_u \subset V$ denote the set of node's u BGP peers, $|W_u| = W \leq M$, and w one of its elements such that $(u,w) \in E$. Let $\varphi_{i,j}(t)$ denote the stability of the route $r_i(t)$ to destination d as received by peering router j ($j \in [1,W]$) at time t. At node u, $r'_{i,stable}(t)=\min\{\varphi_{i,j}(t), \forall j \in [1,W] \mid \{(v_k=u,v_{k-1}=w,\ldots,v_0=v),A\} \in P_{(u,v),d}, \forall w \in W_u\}$ defines −independently of the BGP route selection rules− the selectable route that is the most stable for destination d at time t. Next, we define $\Delta\varphi_i$ as the relative measure of route $r_i$ stability $\varphi_{i,j}$ at time t+1 with respect to stability $\varphi_{i,stable}$ of the most stable route $r'_{i,stable}$ at time t for the same destination d.

```
For i=1 to N /* total number of routes in RT(t+1) */
   if r_i(t+1) is a new route then |Δr_i(t+1)| ← 0
   else if φ_i(t)=0 & φ_i(t+1)=0 then |Δr_i(t+1)| ← 0
       else if φ_i(t+1) > φ_i(t)
           then |Δr_i(t+1)|←[φ_i(t)+1]/[φ_i(t+1)+1]
           else |Δr_i(t+1)|←[φ_i(t)]/[φ_i(t+1)]
           end if
       end if
   end if
end i loop
μ = |ΔRT(t+1)| ← Σ_i Δr_i(t+1)/N
σ² ← Σ_i (Δr_i(t+1) - |ΔRT(t+1)|)² /N
```
**Fig. 2** – *Stability computation for a set of routing entries*

```
For i=1 to N  /* |destinations in Adj_RIB_In| = |Loc_RIB| */
   for j=1 to |W_u| /* number of peers for i^th destination */
       Δφ_{i,j}(t+1)←[φ_{i,j}(t+1)+1]/[φ_{i,stable}(t)+1]
   end j loop
   ΔΦ_i(t+1) ← Σ_j Δφ_{i,j}(t+1)/|W_u|
end i loop
μ = ΔΦ(t+1) ← Σ_i ΔΦ_i(t+1)/N
σ² ← Σ_i (ΔΦ_i(t+1) - ΔΦ(t+1))² /N
```
**Fig. 3** – *Most stable route*

The *best selectable route* from the Adj_RIB_In quantifies the relative stability between incoming routes to the same destination d as learned from all upstream peers and the one amongst them selected by BGP at time t as the best route (thus, following BGP route selection rules). The computational procedure is the same as Fig.3 if one replaces $\varphi_{i,stable}$ by $\varphi_{i,selected}$.

The *differential stability* between the most stable route in the Adj_RIB_In and the selected route stored in the Loc_RIB for the same destination d characterizes the stability of the currently selected routes for a given destination d against most stable routes as learned from upstream neighbors. This metric provides a measure of the stability of the learned routes compared to the stability of the currently selected route. A variant of this metric, denoted $\delta\varphi_i$ ($i \in [1,|D|]$), characterizes the stability of the newly selected path $p^*(u,v)$ at time t for destination d against the stability of the path $p(u,v)$ that is stored as time t in the Loc_RIB for destination d and that would be replaced at time t+1 by the path $p^*(u,v)$: $\delta\varphi_i(t) = \varphi_i(t) - \varphi_i^*(t)$. In turn, if $\delta\varphi_i(t) > 0$, then the replacement of $r_i(t)$ by $r_i^*(t)$ increases stability of the route to destination d; otherwise, the safest decision is to keep the currently selected route $r_i(t)$ stored in the Loc_RIB. Application of the metric $\delta\varphi_i$ during the BGP selection process would prevent replacement of more stable routes by less stable ones but also enable selection of more stable routes than the currently selected routes. However, for this assumption to hold, we must prove the consistency of the stability-based selection with the existing preferential-based route selection model that relies on a path ranking function (i.e., a non-negative, integer-value function $\lambda_u$, defined over $P_{(u,v),d}$, such that if $p_1(u,v)$ and $p_2(u,v) \in P_{(u,v),d}$ and $\lambda_u(p_1) < \lambda_u(p_2)$ then $p_2(u,v)$ is said to be preferred over $p_1(u,v)$). The route selection problem is consistent with the stability function $\delta\varphi(t)$ if $\forall u \in V$ and $p_1(u,v)$ and $p_2(u,v) \in P_{(u,v),d}$ (1) if $\lambda_u(p_1) < \lambda_u(p_2)$ then $\delta\varphi(t) = \varphi_1(t) - \varphi_2(t) \geq 0$ and (2) if $\lambda_u(p_1) = \lambda_u(p_2)$ then $\delta\varphi(t) = 0$. We show in [6] that if $p_1(u,v)$ and $p_2(u,v) \in P_{(u,v),d} \wedge p_2(u,v)$ is embedded in $p_1(u,v)$, then the route selection problem is consistent with the stability function $\delta\varphi$ and the route selection is stretch decreasing.

## 3 Measurement results

This section presents the experimental results obtained by applying the metrics defined in Section 2 to real-world BGP data. The dataset we used was obtained from the Route Views project [5] that comprises archives containing BGP feeds from a set of worldwide distributed Linux PCs running Zebra. As the only policy applied by Route Views sets the next hop to the peer IP address, only Adj_RIBs_In is accessible. As a consequence, the Loc RIB was inferred from the Adj_RIBs_In by implementing a selection process used by Zebra routers.

Fig.4 shows that incoming routes stored in Adj_RIB_In have on average slowly decreasing stability compared to the most stable route (a value close to 1 indicates that incoming routes are nearly as stable as the most stable route). As a result, the plot has a small but positive slope. The average of the maximum metric value per destination d shows a positive but larger slope: the most unstable routes have a faster paced decreasing stability (and spiky pattern confirms their unstable behavior). Further, during the entire observation duration (40 days), a subset of routes continuously presented instabilities leading to a monotonic increase of the metric. It can be seen from Fig.4 that the BGP selected route has on average a better stability than the other routes out of which it is selected (a value close to 1 indicates that incoming routes are nearly as stable as the best selectable route). Comparison between Fig.4 and Fig.6 reveals though that local maxima for the selected route exhibits more spaced and less intensive variations than the most stable route (a lower metric value indicates a higher stability). One can also observe the same monotonously increasing trend of the metric for both the average and the maximum, due to routes with sustained instability. Local maxima in Fig.7 indicate large changes in local route stability, i.e., more routes than the average experienced instabilities but BGP quickly converges to a new stable state since part of the affected routes return to their initial state (thanks to the presence of more stable routes in the Adj_RIB_In, as indicated in Fig.5). Interestingly, Fig.7 shows also that the intensity of the instability increases over time indicating that more routes get affected by the change.

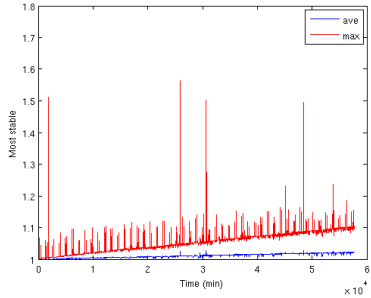 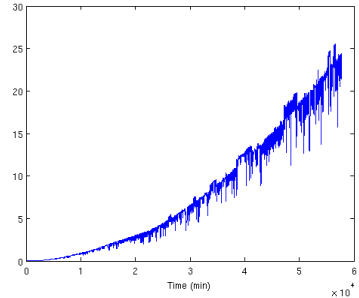 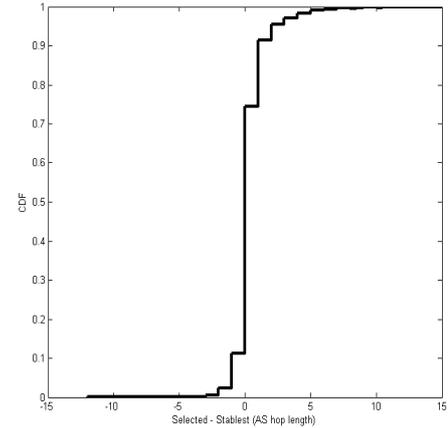

**Fig.4**: Most stable route metric measure  **Fig.5**: Cum.variance against most stable route

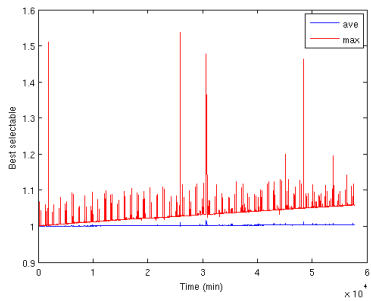 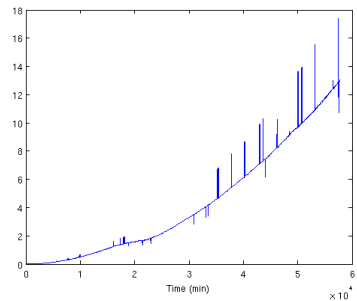

**Fig.6**: Best select. route metric measure  **Fig.7**: Cum.variance against best select. route

**Fig.8**: Number of routes vs diff. in AS_path length

Fig.8 shows the difference between the cumulated percentage of routes against the AS_path length difference between the selected and the most stable route. A positive difference indicates that the replacement of the selected route (using the path ranking function) by the most stable route would decrease the AS_path length compared to the selected route whereas a negative difference indicates that such replacement would increase the AS_path length. From this figure, we can deduce that such replacement would be advisable for about 90% of the selected routes and for 25% percent this replacement would also lead to an AS_path length decrease. Interestingly, only 10% of the routes would be affected by an AS_path length increase if selected based on the stability criteria. Among these 10%, we can also observe from this figure that a significant fraction of the routes is covered if an AS_path length increase of one-hop is acceptable. On the other hand, by admitting a stretch increase corresponding to one additional AS-hop in the AS_path, only a minor fraction of the routes (about 2%) would be penalized by a higher stretch increase (two AS-hops and above). This observation can be seen as the experimental proof that enforcing stability would not come at the detriment of increasing the stretch of most AS-paths.

## 4 Conclusion and perspectives

In this paper, we propose several stability metrics to characterize the local effects of BGP policy- and protocol-induced instabilities on the routing tables. Our experimental results show that the proposed method enables detecting instability events affecting the routing tables, and deriving their impact on the local stability of the routing system. We have also determined a differential stability-based decision criterion that can be taken into account as part of the route selection process. Ongoing work includes verifying the trade-offs between stability-based route selection and the resulting stretch increase/decrease factor on the selected routing paths. Moreover, the relationship between local and global stability will be further elaborated to characterize the effects resulting from the selection of a route that is more stable locally onto the global stability of the routing system, and the model extended to discriminate between protocol- and policy-induced instabilities.

## Références